\documentclass[smallextended]{svmult}

\usepackage{mathptmx}       % selects Times Roman as basic font
\usepackage{helvet}         % selects Helvetica as sans-serif font
\usepackage{courier}        % selects Courier as typewriter font
\usepackage{type1cm}        % activate if the above 3 fonts are
\usepackage{makeidx}         % allows index generation
\usepackage{graphicx}        % standard LaTeX graphics tool
\usepackage{multicol}        % used for the two-column index
\usepackage[bottom]{footmisc}% places footnotes at page bottom

\usepackage{floatrow}

\usepackage[numbers]{natbib}

\makeindex             % used for the subject index
\newcommand{\wrt}{w.r.t. }   % with respect to
\newcommand{\eg}{e.g., }       % for example
\newcommand{\ie}{i.e., }      % that is
\newcommand\etc{etc.}

\newcommand*\samethanks[1][\value{footnote}]{\footnotemark[#1]}

\begin{document}

\title*{Fluid Communities: A Competitive, Scalable and Diverse Community Detection Algorithm}
\titlerunning{Fluid Communities}
%\author{Name of First Author and Name of Second Author}
\author{Ferran Par\'{e}s\thanks{Both authors contributed equally to this work}, Dario Garcia-Gasulla\samethanks, Armand Vilalta, Jonatan Moreno, Eduard Ayguad\'{e}, Jes\'{u}s Labarta, Ulises Cort\'{e}s and Toyotaro Suzumura}
\authorrunning{Par\'{e}s F., Garcia-Gasulla D. et al.}
%\institute{Name of First Author \at Name, Address of Institute, \email{name@email.address}
%\and Name of Second Author \at Name, Address of Institute \email{name@email.address}}
\institute{Ferran Par\'{e}s\samethanks \at Barcelona Supercomputing Center (BSC), Barcelona, Spain, \email{ferran.pares@bsc.es}
\and Dario Garcia-Gasulla\samethanks \at Barcelona Supercomputing Center (BSC), Barcelona, Spain,  \email{dario.garcia@bsc.es}
\and Armand Vilalta \at Barcelona Supercomputing Center (BSC), Barcelona, Spain
\and Jonatan Moreno \at Barcelona Supercomputing Center (BSC), Barcelona, Spain
\and Eduard Ayguad\'{e} \at Barcelona Supercomputing Center (BSC) \& UPC - BarcelonaTECH, Barcelona, Spain
\and Jes\'{u}s Labarta \at Barcelona Supercomputing Center (BSC) \& UPC - BarcelonaTECH, Barcelona, Spain
\and Ulises Cort\'{e}s \at Barcelona Supercomputing Center (BSC) \& UPC - BarcelonaTECH, Barcelona, Spain
\and Toyotaro Suzumura \at Barcelona Supercomputing Center (BSC) \& IBM T.J. Watson, New York, USA
}
\maketitle

\abstract*{We introduce a community detection algorithm (Fluid Communities) based on the idea of fluids interacting in an environment, expanding and contracting as a result of that interaction. Fluid Communities is based on the propagation methodology, which represents the state-of-the-art in terms of computational cost and scalability. While being highly efficient, Fluid Communities is able to find communities in synthetic graphs with an accuracy close to the current best alternatives. Additionally, Fluid Communities is the first propagation-based algorithm capable of identifying a variable number of communities in network. To illustrate the relevance of the algorithm, we evaluate the diversity of the communities found by Fluid Communities, and find them to be significantly different from the ones found by alternative methods.}

\abstract{We introduce a community detection algorithm (Fluid Communities) based on the idea of fluids interacting in an environment, expanding and contracting as a result of that interaction. Fluid Communities is based on the propagation methodology, which represents the state-of-the-art in terms of computational cost and scalability. While being highly efficient, Fluid Communities is able to find communities in synthetic graphs with an accuracy close to the current best alternatives. Additionally, Fluid Communities is the first propagation-based algorithm capable of identifying a variable number of communities in network. To illustrate the relevance of the algorithm, we evaluate the diversity of the communities found by Fluid Communities, and find them to be significantly different from the ones found by alternative methods.}

\section{Introduction} \label{sec:i}

Community detection (CD) extracts structural information of a network unsupervisedly. Communities are typically defined by sets of vertices densely interconnected which are sparsely connected with the rest of the vertices from the graph. Finding communities within a graph helps unveil the internal organization of a graph, and can also be used to characterize the entities that compose it (\eg groups of people with shared interests, products with common properties, \etc).

One of the first CD algorithms proposed in the literature is the Label Propagation Algorithm (LPA) \cite{raghavan2007near}. Although other CD algorithms have been shown to outperform it, LPA remains relevant due to its scalability (with linear computational complexity $O(E)$) and yet competitive results \cite{yang2016comparative}. In this paper we propose a novel CD algorithm: the Fluid Communities (FluidC) algorithm, which also implements the efficient propagation methodology. This algorithm mimics the behaviour of several fluids (\ie communities) expanding and pushing one another in a shared, closed and non-homogeneous environment (\ie a graph), until equilibrium is found. By initializing a different number of fluids in the environment, FluidC can find any number of communities in a graph. To the best of our knowledge, FluidC is the first propagation-based algorithm with this property, which allows the algorithm to provide insights into the graph structure at different levels of granularity. 

%Duh! Who doesnt?
%Finally, FluidC avoids the generation of monster communities (a well known limitation of LPA \cite{leung2009towards}) through the consideration of fluid densities in an intuitive, non-parametric manner.

%This paper is organized as follows: In \S\ref{sec:rw}, we review the related work. The FluidC algorithm and its properties are defined in \S\ref{sec:fca}. In \S\ref{sec:e}, we compare FluidC with top community detection algorithms in terms of clustering performance, while in \S\ref{sec:s} we compare it in terms of computational cost and scalability. \S\ref{sec:r} is dedicated to the reproducibility of our work, providing details on our experimental setting and links to implementations of the algorithm on two different graph processing libraries. Finally, in \S\ref{sec:c} we present our conclusions given the previously reported results.

\section{Related Work} \label{sec:rw}

%Community detection became a popular problem at the beginning of the 21st century, when several algorithms were proposed in a short span of time. 

The most recent evaluation and comparison of CD algorithms was made by \citet{yang2016comparative}, where the following eight algorithms were compared in terms of Normalized Mutual Information (NMI) and computing time: Edge Betweenness \cite{girvan2002community}, Fast greedy \cite{clauset2004finding}, Infomap \cite{rosvall2008maps,rosvall2009map}, Label Propagation \cite{raghavan2007near}, Leading Eigenvector \cite{newman2006finding}, Multilevel (\ie Louvain) \cite{blondel2008fast}, Spinglass \cite{reichardt2006statistical} and Walktrap \cite{pons2005computing}. The performance of these eight algorithms was measured on artificially generated graphs provided by the LFR benchmark \cite{lancichinetti2008benchmark}, which defines a more realistic setting than the alternative GN benchmark \cite{newman2004finding}, including scale-free degree and cluster size distributions. One of the main conclusions of this study is that the Multilevel algorithm is the most competitive overall in terms of CD quality.

%NMI measures the dependence between two variables (in this case the predicted communities and the ground truth communities) taking into account both the quality and the quantity of the predicted communities. 

A similar comparison of CD algorithms was previously reported by \citet{lancichinetti2009community}. In this work twelve algorithms were considered, some of them also present in the study of \cite{yang2016comparative} (Edge Betweenness, Fastgreedy, Multilevel and Infomap). In this study, the algorithms were compared under the GN benchmark, the LFR benchmark, and on random graphs. In their summary, authors recommend using various algorithms when studying the community structure of a graph for obtaining \textit{algorithm-independent information}, and suggest Infomap, Multilevel and the Multiresolution algorithm \cite{ronhovde2009multiresolution} as the best candidates. Results from both \cite{yang2016comparative} and \cite{lancichinetti2009community} indicate that the fastest CD algorithm is the well-known LPA algorithm, due to the efficiency and scalability of the propagation methodology.

\begin{figure}[!t]
    \centering
    \includegraphics[width=\linewidth]{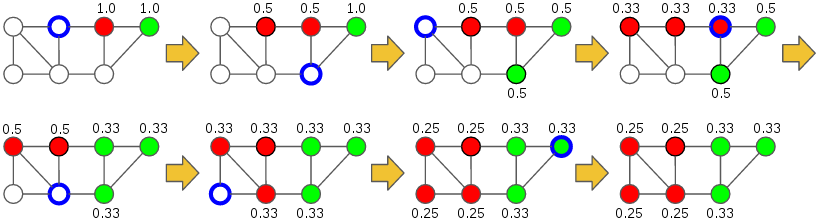}
    \caption{Workflow of FluidC for k=2 communities (red and green). Each vertex assigned to a community is labeled with the density of that community. The update rule is evaluated on each step for the vertex highlighted in blue. The algorithm converges after one complete superstep.}
    \label{fig:fca_workflow}
\end{figure}

\section{Fluid Communities Algorithm}  \label{sec:fca}

%The Fluid Communities (FluidC) algorithm is a community detection algorithm based on the idea of introducing a number of fluids (\ie communities) within a non-homogeneous environment (\ie a non-complete graph), where fluids will expand and push each other influenced by the topology of the environment until a stable state is reached. As a fluid community spreads through more vertices its density decreases, which reduces its strength to conquer and defend vertices. Unlike the hop attenuation of the LPA variants, densities in FluidC depend exclusively on the community size, and may increase or decrease regardless of the initialization setup.

The Fluid Communities (FluidC) algorithm is a CD algorithm based on the idea of introducing a number of fluids (\ie communities) within a non-homogeneous environment (\ie a non-complete graph), where fluids will expand and push each other influenced by the topology of the environment until a stable state is reached.

Given a graph $G=(V,E)$ composed by a set of vertices $V$ and a set of edges $E$, FluidC initializes $k$ fluid communities $\mathcal{C} = \{c_{1}..c_{k}\}$, where $0<k\leq|V|$. Each community $c\in\mathcal{C}$ is initialized in a different and random vertex $v \in V$. Each initialized community has an associated density $d$ within the range $(0,1]$. The density of a community is the inverse of the number of vertices composing said community:

\begin{equation}
    \label{eq:density}
    d(c) = \frac{1}{\left | v \in c \right |}
\end{equation}

Notice that a fluid community composed by a single vertex (\eg every community at initialization) has the maximum possible density ($d=1.0$).

FluidC operates through supersteps. On each superstep, the algorithm iterates over all vertices of $V$ in random order, updating the community each vertex belongs to using an update rule. When the assignment of vertices to communities does not change on two consecutive supersteps, the algorithm has converged and ends.

The update rule for a specific vertex $v$ returns the community or communities with maximum aggregated density within the ego network of $v$. The update rule is formally defined in Equations \ref{update_rule_1} and \ref{update_rule_2}.

% consist on multiplying community wise the densities by the number of vertices belonging to each fluid community among neighbours of $v$ and itself. Formally, we define the updating rule as follows

\begin{equation}\label{update_rule_1}
    \mathcal{C}_{v}^{'} = {argmax}_{c \in \mathcal{C}} \sum_{ w \in \{v, \Gamma(v)\}} d(c) \times \delta(c(w),c)
\end{equation}

\begin{equation}\label{update_rule_2}
    \delta(c(w),c) = \left\{\begin{array}{@{}ccc@{}}
    1 & ,if & c(w) = c\\ 
    0 & ,if & c(w) \neq c 
    \end{array}\right.
\end{equation}

where $v$ is the vertex being updated, $\mathcal{C}^{'}_{v}$ is the set of candidates to be the new community of $v$, $\Gamma(v)$ are the neighbours of $v$, $d(c)$ is the density of community $c$, $c(w)$ is the community vertex $w$ belongs to and $\delta(c(w),c)$ is the Kronecker delta. 

%If $\mathcal{C}_{v}^{'}$ is composed by a single community $c$, the update rule sets $c$ as the new community of $v$. If $\mathcal{C}_{v}^{'}$ is composed by two or more communities with equal density, but the current community of the vertex $v$ is not among those, the update rule chooses a random community within $\mathcal{C}_{v}^{'}$ as the new community of $v$. If $\mathcal{C}_{v}^{'}$ is composed by two or more communities with equal density and the current community of the vertex $v$ is among those, $v$ does not change its community. This last rule guarantees that no community will ever be eliminated from the graph, since, when a community $c$ is compressed into a single vertex $v$, $c$ has the maximum possible density on the update rule of $v$ (\ie $1.0$) which guarantees that $c\in\mathcal{C}_{v}^{'}$. An example of the FluidC algorithm behaviour is shown in Figure \ref{fig:fca_workflow}. 

Notice that $\mathcal{C}_{v}^{'}$ could contain several community candidates, each of them having equal maximum sum. If $\mathcal{C}_{v}^{'}$ contains the current community of the vertex $v$, $v$ does not change its community. However, if $\mathcal{C}_{v}^{'}$ does not contain the current community of $v$, the update rule chooses a random community within $\mathcal{C}_{v}^{'}$ as the new community of $v$. This completes the formalization of the update rule:

\begin{equation}\label{update_rule_3}
    c^{'}(v) = \left\{\begin{array}{@{}ccc@{}}
    x \sim \mathcal{U}(\mathcal{C}^{'}_{v}) & ,if & c(v) \notin \mathcal{C}^{'}_{v}\\
    c(v) & ,if & c(v) \in \mathcal{C}^{'}_{v}\\
    \end{array}\right.
\end{equation}
where $c^{'}(v)$ is the community of vertex $v$ at the next superstep, $\mathcal{C}_{v}^{'}$ is the set of candidate communities from equation \ref{update_rule_1} and $x \sim \mathcal{U}(\mathcal{C}_{v}^{'})$ is the random sampling from a discrete uniform distribution of the $\mathcal{C}_{v}^{'}$ set.

Equation \ref{update_rule_3} guarantees that no community will ever be eliminated from the graph since, when a community $c$ is compressed into a single vertex $v$, $c$ has the maximum possible density on the update rule of $v$ (\ie $1.0$) guaranteeing $c\in\mathcal{C}_{v}^{'}$, and thus $c^{'}(v)=c$. An example of FluidC algorithm behaviour is shown in Figure \ref{fig:fca_workflow}.

%Equation \ref{update_rule_3} guarantees that no community will ever be eliminated from the graph and the reason why is explained in two steps. Firstly, if a community $c$ losses almost all its vertices and ends up into the extreme case of having only one last vertex $v$, when we evaluate next community of vertex $v$, $c$ would obtain the maximum possible sum value (\ie 1.0), guaranteeing that $c$ would be in the set of candidates $\mathcal{C}_{v}^{'}$. However, it is not guaranteed for $c$ to be the only community candidate, since it could be other communities that obtain also a maximum amount of $1.0$ at Equation \ref{update_rule_1}. So, as second step, if $c\in\mathcal{C}_{v}^{'}$ we avoid random selection at Equation \ref{update_rule_3}, obtaining the same $c$ community from previous superstep at this extreme case. An example of FluidC algorithm behaviour is shown in Figure \ref{fig:fca_workflow}.

\subsection{Properties}

\textit{FluidC is asynchronous}, where each vertex update is computed using the latest partial state of the graph (some vertices may have updated their label in the current superstep and some may not). A straight-forward synchronous version of FluidC (\ie one where all vertex update rules are computed using the final state of the previous superstep) would not guarantee that densities are consistent with Equation \ref{eq:density} at all times (\eg a community may lose a vertex but its density may not be increased immediately in accordance). Consequently, a community could lose all its vertices and be removed from the graph.

FluidC \textit{allows for the definition of the number of communities to be found}, simply by initializing a different number of fluids in the graph. This is a desirable property for data analytics, as it enables the study of the graph and its entities at several levels of granularity. Although a few CD algorithms already had this feature (\eg Walktrap, Fastgreedy), none of those were based on the efficient propagation method.

%\begin{wrapfigure}[13]{R}{0.5\textwidth}

\begin{figure}[t]
    \floatbox[{\capbeside\thisfloatsetup{capbesideposition={left,center},capbesidewidth=4.5cm}}]{figure}[\FBwidth]
    {\centering
    \includegraphics[scale=.20]{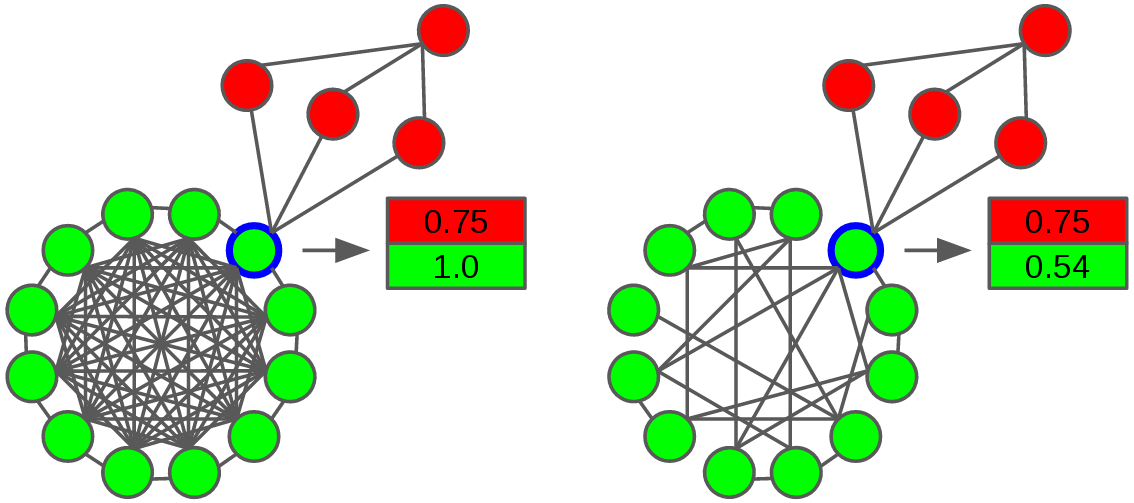}}
    {\caption{Two cases of update rule on a vertex (highlighted in blue). Left case shows a densely connected green community which can successfully defend the vertex. In the right case, the green community is sparser and will lose the highlighted vertex (but only this one) to the red community. Numbers correspond to sums from Equation \ref{update_rule_1}.}}
    \label{fig:fca_big_coms}
\end{figure}

%%%%%%%%%%%%%%%%%%%%%%%%%%%%%%%%%%%%%%%
%%%%%%%%%%%%%%%%%%%%%%%%%%%%%%%%%%%%%%%
Another interesting feature of FluidC is that it \textit{avoids the creation of monster communities} in a non-parametric manner. Due to the spread of density among the vertices that compose a community, a large community (when compared to the rest of communities in the graph) will only be able to keep its size and expand by having a favourable topology (\ie having lots of intra-community edges which make up for its lower density). Figure \ref{fig:fca_big_coms} shows two cases of this behaviour, one where a large community is able to defend against external attack, and one where it is not.
%%%%%%%%%%%%%%%%%%%%%%%%%%%%%%%%%%%%%%%
%%%%%%%%%%%%%%%%%%%%%%%%%%%%%%%%%%%%%%%

\textit{FluidC is designed for connected, undirected, unweighted graphs}, and variants of FluidC for directed and/or weighted graphs remain as future work. However, FluidC can be easily applied to a disconnected graph $G^{'}$ just by performing an independent execution of FluidC on each connected component of $G^{'}$ and appending the results.

% \textbf{Statement 1} \newline
% Given a connected graph $G=(V,E)$, and a set of communities $\mathcal{C}$ such that $|\mathcal{C}|=k$
% \begin{enumerate}
%     \item $\forall v\in c,\forall c\in \mathcal{C}$ at superstep $t=i$, then $\exists c^{'}\in \mathcal{C}$ such that $v\in c^{'}$ at $t=i+1$\newline
%     (A vertex assigned to a community in a superstep will have an assigned community in all subsequent supersteps)
%     \item $\forall v\in \Gamma(w),\forall w\in c, \forall c\in\mathcal{C}$ at $t=i$, then $\exists c^{'}\in \mathcal{C}$ such that $v\in c^{'}$ at $t=i+1$\newline
%      (A vertex neighbour of an assigned vertex in a superstep will have an assigned community on the next superstep)
% \end{enumerate}
% Since the algorithm does not converge while the assignment of vertices to communities change between consecutive supersteps, these two points guarantee that by the time the algorithm converges all vertices are assigned to a community.

% \textbf{Statement 2} \newline
% %Given a connected graph $G=(V,E)$ where connected ensures $|\Gamma(v)|>0 \ \forall v \in V$, and given a $k \in \N$ conditioned to $0<k\leq|V|$: \newline
% Given a connected graph $G=(V,E)$, and a set of communities $\mathcal{C}$ such that $|\mathcal{C}|=k, \forall k>1$
% \begin{enumerate}
%     \item $\forall c \in \mathcal{C}, |c|=1$ at step $t^{'}=j$, $\exists c, |c|\geq1$ at step $t^{'}=j+1$.
% \end{enumerate}
% This point prove that any community $c \in \mathcal{C}$ will not disappear during FluidC execution.

\section{Evaluation} \label{sec:e}

To evaluate performance we use the LFR benchmark \cite{lancichinetti2008benchmark}, measuring NMI obtained on a set of graphs with six different graph sizes ($|V|=$ 233, 482, 1000, 3583, 8916 and 22186) and 25 different mixing parameters ($\mu$ from 0.03 to 0.75). The mixing parameter is the average fraction of vertex edges which connect to vertices from other communities \cite{lancichinetti2008benchmark}.

\begin{figure}[t]
    \renewcommand{\arraystretch}{0}
    \centering
    \begin{tabular}{cc}
        (a) Fastgreedy & (b) Infomap \\
        \includegraphics[width=0.46\linewidth]{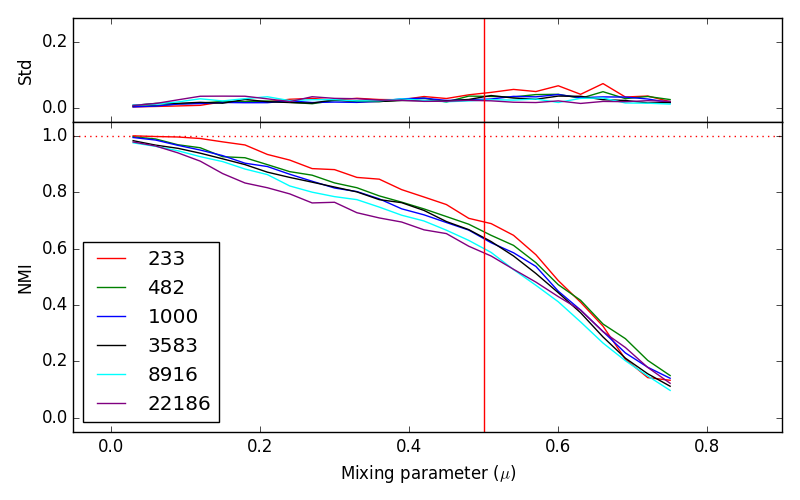} &
        \includegraphics[width=0.46\linewidth]{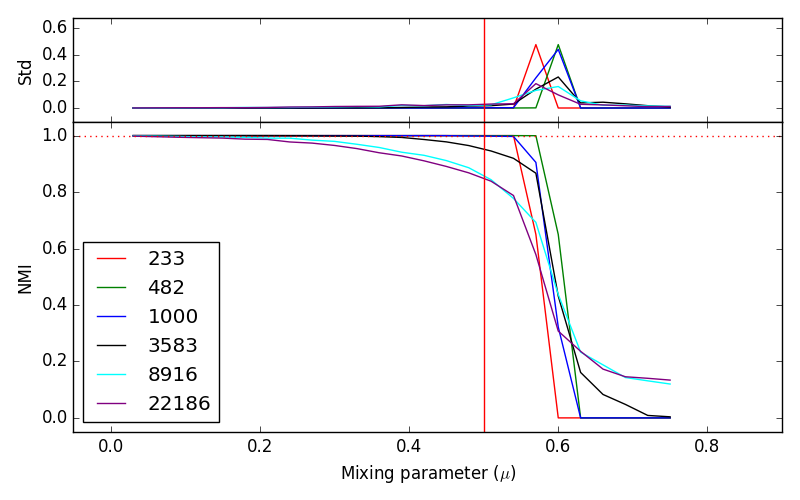}\\
        (c) Multilevel & (d) Walktrap \\
        \includegraphics[width=0.46\linewidth]{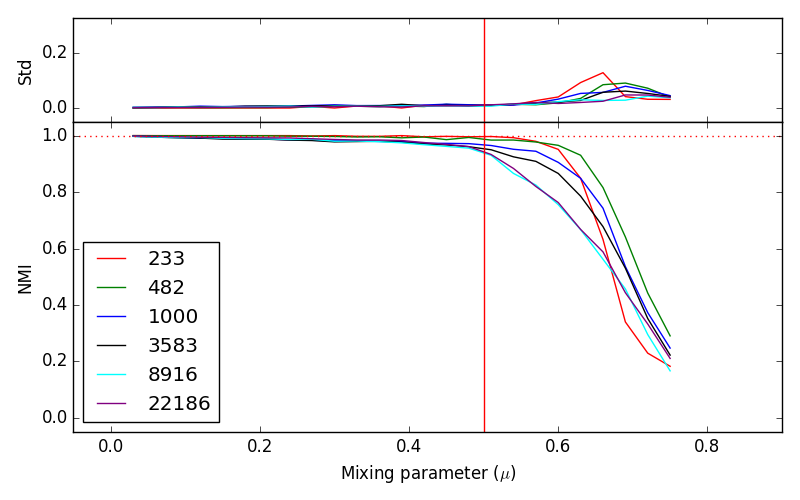} &
        \includegraphics[width=0.46\linewidth]{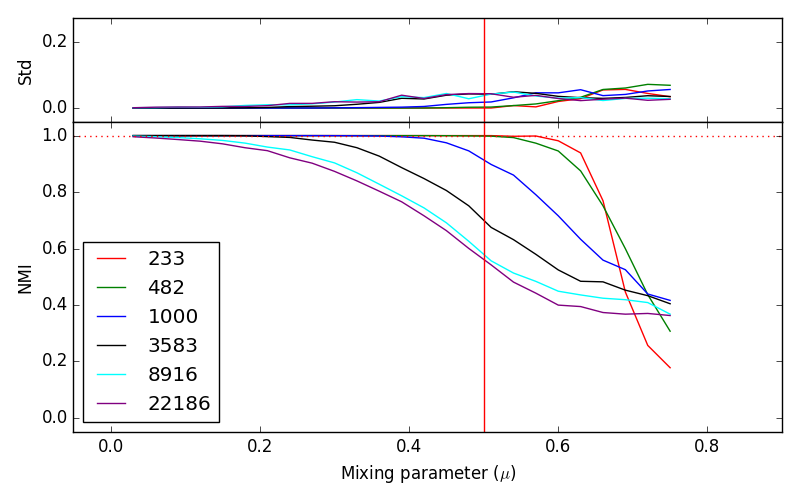}\\
        (e) Label Propagation & (f) Fluid Communities \\
        \includegraphics[width=0.46\linewidth]{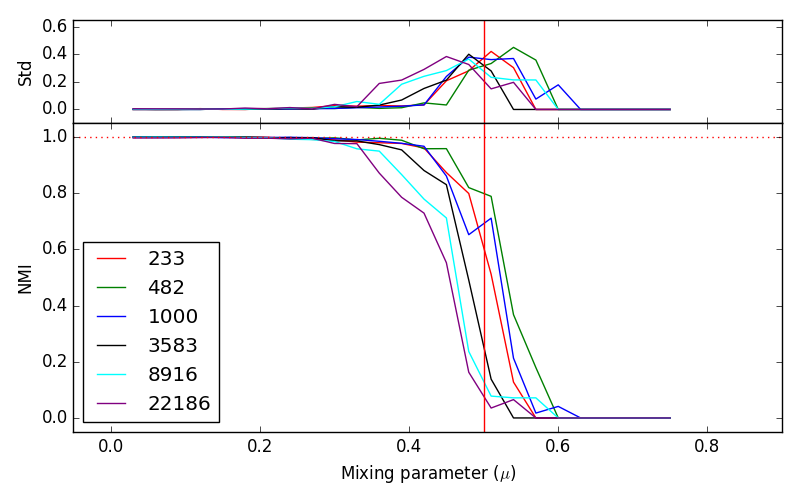} &
        \includegraphics[width=0.46\linewidth]{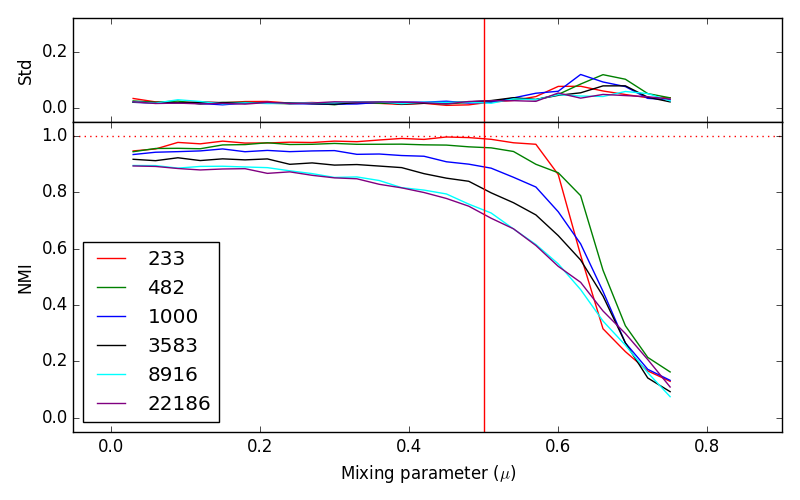}\\
    \end{tabular}
    \caption{NMI performance of six CD algorithms. Each Panel is divided in two plots, the bottom one shows the average NMI performance over 20 random graphs generated with the same properties, including size and mixing parameter. The top plot shows the standard deviation. Different plotted lines correspond to different graph sizes. For reference, each panel contains two lines, one vertical to mark the 0.5 mixing parameter ($\mu$) and one dotted horizontal line to mark the perfect NMI score (NMI $=1.0$). Walktrap performance reported here differs significantly from \citet{yang2016comparative} due to an error in their experiments, acknowledged by the authors.}
    \label{fig:lfr_bench}
\end{figure}

To guarantee consistency, 20 different graphs were generated for each combination of graph size and mixing parameter. This results in a total of 3,000 different graphs (6 graph sizes $\times$ 25 mixing parameter values $\times$ 20 graphs), and is the same evaluation strategy used in \cite{yang2016comparative}. Besides the graph size and mixing parameter, the LFR benchmark also requires a list of hyperparameters to generate a graph. We use the same ones defined in \cite{yang2016comparative}; maximum degree $0.1\times|V|$, maximum community size $0.1\times|V|$, average degree $20$, degree distribution exponent $-2$ and community size distribution exponent $-1$.

These experiments were performed on the five most competitive CD algorithms evaluated in \cite{yang2016comparative} (Fastgreedy, Infomap, Label Propagation, Multilevel and Walktrap) and FluidC. We report the results in Figure \ref{fig:lfr_bench}, where each panel contains two plots. The bottom one shows average performance in NMI on the 20 graphs built for the various values of $\mu$ (shown on the horizontal axis), while the top one shows the corresponding standard deviation (Std). Each plot line represents the results of an algorithm on a different graph size (see panel legend). Among all normalization variants of the Mutual Information metric we use the geometric normalization, \ie dividing by square root of both entropies.

Before analyzing the results, let us clarify two aspects of the evaluation process. First, the LFR benchmark may generate disjoint graphs. When this is the case, an independent execution of FluidC is computed on each connected subgraph separately, and the communities found on the different subgraphs are appended to measure the overall NMI. And second, FluidC requires to specify the number of communities to be found $k$, which is an unknown parameter. For comparability reasons, we report the results obtained using the $k$ resulting in highest modularity. This is analogous to what other algorithms which also require $k$ do (\eg Fastgreedy and Walktrap).

In our experiments, Multilevel achieved top NMI results on most generated graphs, while Fastgreedy and LPA were clearly inferior to the rest of algorithms. The remaining three algorithms, Walktrap, Infomap and FluidC, were competitive, and had results close to Multilevel. In the context of Walktrap and Infomap, FluidC has a rather particular behaviour. It is better on large graph sizes than Walktrap; for the largest graph computed ($|V|=22,186$), FluidC outperforms Walktrap for all $\mu$ values in the range $[0.33,0.66]$. FluidC is also more resistant to large mixing parameters than Infomap, which is unable to detect relevant communities for $\mu>0.55$. 

The performance of FluidC is slightly sub-optimal (NMI between 0.9 and 0.95) on low mixing parameters ($\mu\leq0.4$). This is because communities generated in a graph with low mixing parameters are very densely connected, and only have a few edges connecting them with other communities, edges that act as bottlenecks. These bottlenecks can sometimes prevent the proper flow of communities in FluidC, which leads to sub-optimal results. 

In practice, the sub-optimal performance of FluidC on graphs with very low mixing parameter is a minor inconvenience. Real world graphs are often large and have relatively high mixing parameters, a setting where FluidC is particularly competent. In contrast, the recommended algorithm for processing graphs with low mixing parameters would be LPA, as it finds the optimal result in these cases, and it is faster and scales better than the alternatives.

\section{Scalability} \label{sec:s}

% \begin{wrapfigure}[15]{r}{0.5\textwidth}
%     \renewcommand{\arraystretch}{0}
%     \centering
%     \includegraphics[width=\linewidth]{FluidC_vs_LPA_mean_iter_times_plot}
%     \caption{Computing time per iteration for the FluidC and LPA algorithms, when processing graphs of varying size and mixing parameter.}
%     \label{fig:iter_times}
% \end{wrapfigure}

The main purpose of the FluidC algorithm is to provide high quality communities in a scalable manner, so that good quality communities can also be obtained from large scale graphs. In the previous section we saw how the performance of FluidC in terms of NMI is comparable to the best algorithms in the state-of-the-art (\eg Multilevel, Walktrap and Infomap). Next we evaluate FluidC scalability, to show its relevance in the context of large networks.

\begin{figure}[!t]
    \renewcommand{\arraystretch}{0}
    \centering
    \begin{minipage}[t]{0.47\textwidth}
        \includegraphics[width=\linewidth]{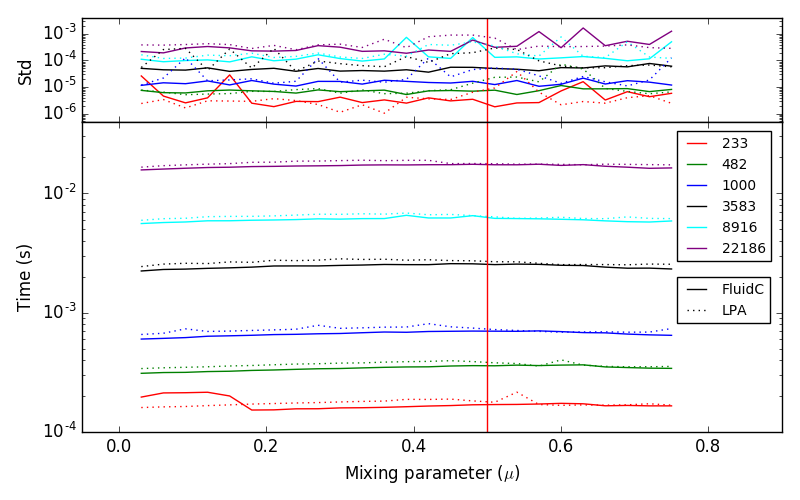}
        %\caption{Computing time per iteration for the FluidC and LPA algorithms, when processing graphs of varying size and mixing parameter.}
        %\label{fig:iter_times}
        %\vspace{16pt}
    \end{minipage}
    \begin{minipage}[t]{0.47\textwidth}
        \includegraphics[width=\linewidth]{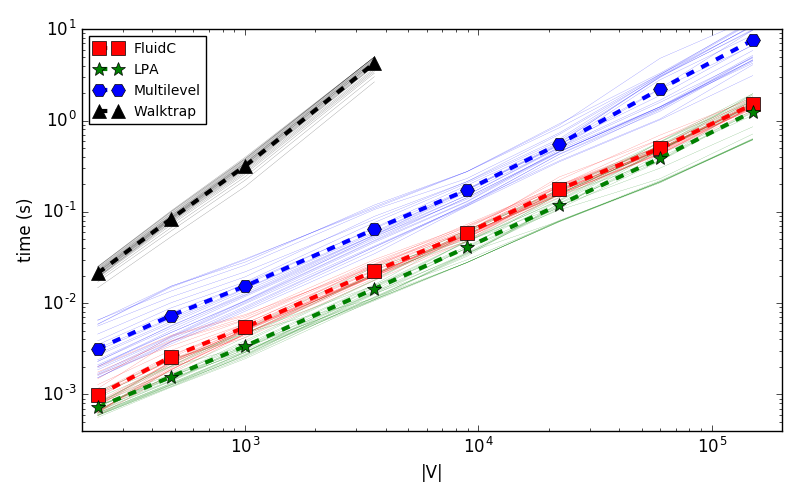}
        %\caption{Scalability of FluidC, LPA, Multilevel and Walktrap on graphs up to 150,000 vertices. Dashed line shows the average over the various mixing parameters values (from 0.03 to 0.75).}
        %\label{fig:top_times}
    \end{minipage}
    \vspace{-10pt}
    \caption{Left: Computing time per iteration for the FluidC and LPA algorithms, when processing graphs of varying size and mixing parameter. Right: Scalability of FluidC, LPA, Multilevel and Walktrap on graphs up to 150,000 vertices. Dashed line shows the average over the various mixing parameters values (from 0.03 to 0.75).}
    \label{fig:times}
\end{figure}

To analyze the computational cost of FluidC we first compare the cost of one full superstep (checking and updating the communities of all vertices in the graph) with that of LPA. LPA is the fastest and more scalable algorithm in the state-of-the-art \cite{yang2016comparative}, which is why we use it as baseline for scalability along this section. Figure \ref{fig:times} shows the average time per iteration, using the same type of plots used in the NMI evaluation. Results indicate that the computing time per iteration of FluidC is virtually identical to that of LPA for all graph sizes and mixing parameters. Significantly, both algorithms are almost unaffected by a varying mixing parameter.

\begin{figure}[!t]
    \renewcommand{\arraystretch}{0}
    \centering
    \begin{tabular}{cc}
        Fluid Communities & Label Propagation\\
        \includegraphics[width=0.47\linewidth]{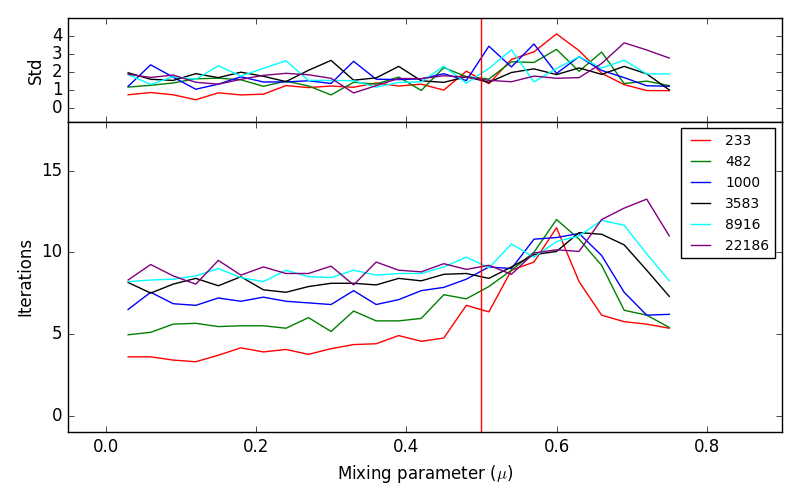} & \includegraphics[width=0.47\linewidth]{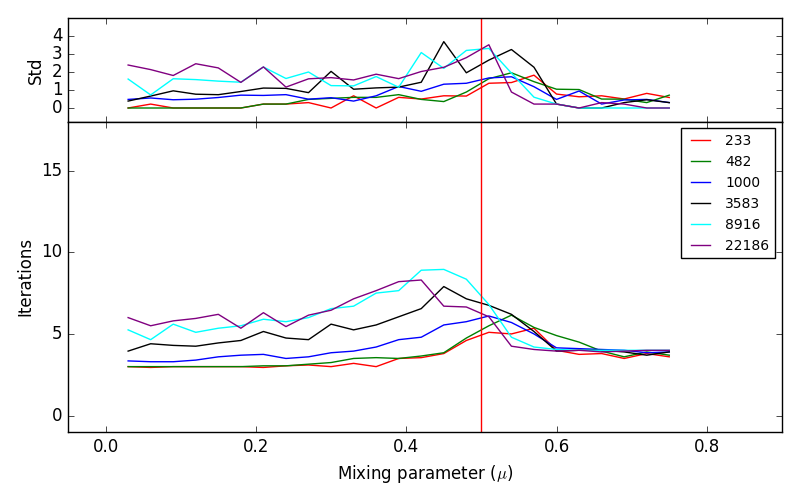}
    \end{tabular}
    \vspace{-10pt}
    \caption{Number of iterations until convergence for the FluidC and LPA algorithms, when processing graphs of varying size and mixing parameter.}
    \label{fig:iter}
\end{figure}

Beyond the cost of a single superstep, we also explore the total number of supersteps needed for the algorithm to converge. Figure \ref{fig:iter} shows that information for both FluidC and LPA. For this experiment we set the FluidC parameter $k$ to the ground truth. This comparison is relevant for mixing parameters up to 0.5. Beyond that value, LPA produces a single monster community after three supersteps (NMI = 0.0, see Figure \ref{fig:lfr_bench}, Panel e). Nevertheless, the number of supersteps needed by FluidC to converge is never above 13.

% In the case of LPA, the number of iterations is rather stable for low mixing parameters ($\mu < 0.3$), and it slightly increases starting on $\mu$ values around 0.45. At $\mu > 0.6$ the number of iterations of LPA coverges to 4, probably because at this point LPA is unable to produce relevant communities (NMI = 0.0).

% The number of iterations of FluidC can be divided in three phases based on the mixing parameter $\mu$ as follows:
% \begin{enumerate}
%     \item When the mixing parameter is low ($\mu < 0.4$), the number of iterations is stable for each graph size, and slightly higher than LPA (less than twice as many iterations).
%     \item When the mixing parameter is close to $\mu = 0.6$, all graph sizes converge on 10 iterations. At this point FluidC is largely outperforming LPA in terms of NMI, only by doing a few more iterations.
%     \item When the mixing parameter is high ($\mu > 0.6$), the number of iterations becomes varied again, and is correlated with the graph size.
% \end{enumerate}

% \begin{tabular}
%     \begin{table}[]
%         \centering
%         \begin{tabular}{cc}
%             \begin{figure}
%                 \renewcommand{\arraystretch}{0}
%                 \centering
%                 \includegraphics[width=0.5\linewidth]{TOP_times}
%                 \caption{Scalability of FluidC, LPA, Multilevel and Walktrap on graphs up to 150,000 vertices. Dashed line shows the average over the various mixing parameters values (from 0.03 to 0.75).}
%                 \label{fig:top_times}
%             \end{figure} & 
%             aloha\\
%         \end{tabular}
%     \end{table}
% \end{tabular}

These results indicate that FluidC and LPA are similar both in time per superstep and number of supersteps, which implies that both algorithms are analogous in terms of computational cost (with linear complexity $O(E)$) and scalability. To provide further evidence in that regard, and to also evaluate the scalability of the most relevant alternatives, next we consider the evaluation of larger graphs. We generate graphs with 60,000 and 150,000 vertices following the same methodology described in \S\ref{sec:e}, and measure the computing times of LPA, FluidC, Multilevel and Walktrap (the three fastest algorithms according to \cite{yang2016comparative} plus FluidC). Figure \ref{fig:times} shows the scalability of each algorithm, where the continuous lines in the background correspond to different mixing parameters (from 0.03 to 0.75), and the big dashed line with markers indicates the computed mean over all the 25 mixing parameters. 

According to the results shown in Figure \ref{fig:times}, LPA has the lowest computing time, closely followed by FluidC. However, LPA results are mistakenly optimistic, since the algorithm is particularly fast for large mixing parameters, where it obtains zero NMI after doing only three supersteps (see Panel e of Figure \ref{fig:lfr_bench}, and Figure \ref{fig:iter}). If this is taken into account, LPA and FluidC have an analogous scalability.

Walktrap is considerably slower than the rest, and results for graphs larger than 3,000 vertices are not shown. Multilevel is 5x slower than LPA/FluidC, and its cost grows faster. While the slope of LPA/FluidC computed through a linear regression is roughly $10^{-6}$, the slope of Multilevel is close to $10^{-5}$. Processing a large scale graph like PageGraph \cite{meusel2014graph} ($|V|=3,500M$ vertices) would take approximately 9 hours for FluidC while more than two days (approximately 49 hours) for Multilevel.

%The slope of Walktrap on the other hand is around $10^{-3}$.

%According to \cite{yang2016comparative}, the computational cost and scalability of the rest of algorithms evaluated in Figure \ref{fig:lfr_bench} is equal or worse than Walktrap.

% boldi2004webgraph, meusel2014graph

\section{Diversity of Communities} \label{sec:div}

%%%%%%%%%%%%%%%%%%%%%%%%%%%%%%%%%%%%%%%
%%%%%%%%%%%%%%%%%%%%%%%%%%%%%%%%%%%%%%%
%The evaluation of unsupervised learning methods, such as community detection algorithms, is limited by the inherent lack of a universal ground truth. Since the No Free Lunch theorem \cite{wolpert1996lack} also applies to the community detection problem \cite{peel2016ground}, one cannot claim that an algorithm A universally outperforms another algorithm B, as there are as many cases where algorithm A will outperform B as cases where algorithm B will outperform A. 
%%%%%%%%%%%%%%%%%%%%%%%%%%%%%%%%%%%%%%%
%%%%%%%%%%%%%%%%%%%%%%%%%%%%%%%%%%%%%%%

Synthetic graphs generated by benchmarks like LFR can be used to evaluate the ability of algorithms at finding the community structures pre-defined by those benchmarks. This kind of results are useful for understanding the strengths and weaknesses of algorithms \wrt certain structural properties (\eg $\mu$ and graph size). However, graphs obtained from real world data will rarely contain a single community structure, as complex data can be typically sorted in several coherent but unrelated ways (\eg group products by sell volume or by sell dates). For this reason it is recommended to use several CD algorithms when analyzing a given graph, to obtain a variety of algorithm independent information \cite{lancichinetti2009community}. In this context, the relevance of a CD algorithm is also affected by how diverse are the communities it is capable of finding, in comparison with the communities that the rest of the algorithms find.

% In our experiments we evaluated six algorithms, which can be categorized in three communities: Modularity based algorithms (Multilevel and Fastgreedy) look for communities which optimize their internal connectivity with reference to a randomized null model with the same degree distribution \cite{orman2011accuracy}. Random-walk algorithms (Walktrap and Infomap) partition the network based on the behaviour of a random walker. Finally, node neighbourhood algorithms (LPA and FluidC) simulate a diffusion over the network topology.

% We argue that, in practice, using at least one algorithm of each family is recommended to get the most rich insight into the community structure of a given graph. To illustrate the relevance of this recommendation, we build a graph using LFR generation procedure with different ground truths, and analyze how different algorithms unconver different ground truths.

To evaluate the relevance of FluidC in terms of diversity, we execute the six previously evaluated algorithms on a set of graphs with multiple ground truths. If two algorithms consistently find different ground truths from the ones available within multi-ground truth graphs, it can be argued that both may provide different insights into the community structure of a graph. We build a total of 30 multi-ground truth graph, each of them containing two independent ground truths, each ground truth composed by four communities. Our multi-ground truth graphs are obtained by first generating two independent graphs (of size $|V|=22,186$) with four disconnected communities ($\mu=0$) and then appending the edges of both. The rest of the LFR parameters used are the same ones used for evaluation at \S\ref{sec:e}, except for minimum and maximum community size which are $0.2|V|$ and $0.3|V|$ respectively.

Given a multi-ground truth graph $G$ with two ground truths $T_{1},T_{2}$, we categorize a CD algorithm $\mathcal{A}$ in one of three values of $\theta$; the algorithm finds $T_{1}$ over $T_{2}$ ($\theta=1$), the algorithm finds $T_{2}$ over $T_{1}$ ($\theta=-1$), the algorithm finds $T_{1}$ and $T_{2}$ to the same degree ($\theta=0$). $\mathcal{A}$ finds ground truth $T_{x}$ over ground truth $T_{y}$ when NMI($\mathcal{A}$,$T_{x}$) - NMI($\mathcal{A}$,$T_{y}$) $>$ $\alpha$, where $\alpha$ is a predefined threshold.

%the difference between the NMI score of $\mathcal{A}$ \wrt$T_{x}$ and the NMI score of $\mathcal{A}$ \wrt $T_{y}$ is larger than a given threshold $\alpha$.

% Si hi ha espai el posem, si no el paragraf anterior ja es prou formal
% Si es posa cal adaptar la nomeclatura: alpha, T_{}
% Formally:
% \begin{equation}
%     \label{eq:beh}
%     \theta = \left\{\begin{array}{@{}ccc@{}}
%     -1 & ,if & (NMI_{1}-NMI_{2})<-0.5\\
%     0  & ,if & -0.5<(NMI_{1}-NMI_{2})<0.5\\
%     1  & ,if & 0.5<(NMI_{1}-NMI_{2})\\
%     \end{array}\right.
% \end{equation}

\begin{figure}[!t]
    \renewcommand{\arraystretch}{0}
    \centering
    \begin{tabular}{cc}
        \hspace{-10pt}
        \includegraphics[width=0.52\linewidth]{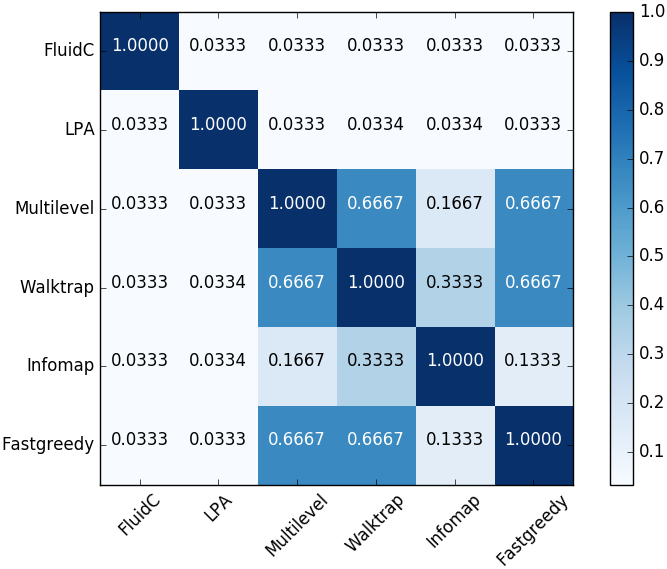} &
        \hspace{-25pt}
        \includegraphics[width=0.52\linewidth]{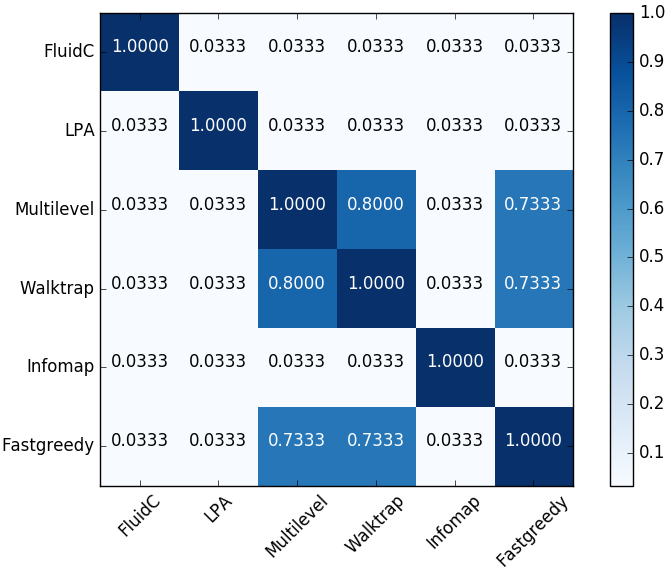}\\
    \end{tabular}
    \vspace{-15pt}
    \caption{Average similarity matrix with $\alpha=0.5$ (left) and $\alpha=0$ (right) over 30 multi-ground truth graphs. These matrices indicate a qualitative degree of similarity among algorithm behaviour.}
    \label{fig:div}
\end{figure}

After one hundred executions, the behaviour of an algorithm \wrt a multi-ground truth graph is represented by one hundred $\theta$ values. A similarity between the behaviour of two algorithms on the same graph can be computed by applying the Chi-square test on the two corresponding series of $\theta$ values, resulting in a probability of both series having the same categorical distribution. Qualitatively, that probability refers to how likely both algorithms are of having the same behaviour. We built 30 similarity matrices, each one corresponding to a multi-ground truth graph. For visualization purposes, we average them in Figure \ref{fig:div}.

%To continue here
% We show one similarity matrix using $\alpha=0$, and one using $\alpha=0.5$. The main reason to use a couple of $\alpha$ values is 

The resultant similarity matrix provides qualitative insight on the diversity of the algorithms under consideration. To validate the consistency of the approach \wrt the $\alpha$ threshold, we report the similarity matrix corresponding to setting $\alpha=0.5$ and $\alpha=0$. Since the NMI score is in the range $[0,1]$, $\alpha=0.5$ implies that an algorithm finds one ground truth over the alternative at least by doubling its NMI score. On the other hand, the more permissive $\alpha=0$ considers that an algorithm finds one ground truth over the alternative just by obtaining a larger NMI. In this case, $\theta=0$ values are rare, as that would require both NMI to be identical. Additional values of $\alpha$ within the range $[0,0.5]$ were analyzed, and the corresponding similarity matrices where found to be consistent with the two reported ones.

The similarity matrices of Figure \ref{fig:div} indicate that Multilevel, Walktrap and Fastgreedy have similar behaviours on multi-ground truth graphs, as all three algorithms result in very similar $\theta$ values. Significantly, this relation is stronger when $\alpha=0$, which indicates that all three algorithms consistently prioritize the same ground truth over the alternative. Notice all three algorithms use hierarchical bottom-up agglomeration, while Fastgreedy and Multilevel are modularity-based methods \cite{fortunato2010community}.

The remaining three algorithms, Infomap, LPA and FluidC, show a much more diverse behaviour. The pattern with which each of these algorithms finds one ground truth over the alternative differs significantly from the other five. These results indicate that the proposed FluidC algorithm provides significantly diverse communities when compared to the algorithmic alternatives. In this context, FluidC may contribute to provide \textit{algorithm-independent information} by being computed alongside Multilevel and Infomap on the same graph, all of which have a competitive performance according to the LFR benchmark as shown in \S\ref{sec:e}.

\section{Reproducibility} \label{sec:r}

All the experiments presented in this paper have been computed on a computer with OpenSUSE Leap 42.2 OS (64-bits), with an Intel(R) Core(TM) i7-5600U CPU @ 2.60GHz and 16GB of DDR3 SDRAM. An open source implementation of the FluidC algorithm has been made publicly available to the community at Github (\url{github.com/FerranPares/Fluid-Communities}). It has been integrated into the \texttt{networkx} (\url{github.com/networkx}) and \texttt{igraph} (\url{github.com/igraph}) graph libraries. For consistency, all scalability experiments were performed using the \texttt{igraph} graph library.

\section{Conclusions} \label{sec:c}

In this paper we propose a novel CD algorithm called Fluid Communities (FluidC). Through the well established LFR benchmark we demonstrate that FluidC identifies high quality communities (measured in NMI, see Figure \ref{fig:lfr_bench}), getting close to the current best alternatives in the state-of-the-art (\eg Multilevel, Walktrap and Infomap). In this context, FluidC is particularly competent on large graphs and on graphs with large mixing parameters. The main limitation of FluidC in terms of NMI performance is that it does not fully recover the ground truth communities on graphs with small mixing parameters due to the effect of bottleneck edges. However, at larger mixing parameters (a more realistic environment) FluidC gets competitive to state-of-the-art algorithms. Although FluidC does not clearly outperform the current state-of-the-art in terms of NMI on the LFR benchmark, the importance of the contribution can be summarized both in terms of scalability and diversity.

In terms of scalability, FluidC, together with LPA, represents the state-of-the-art in CD algorithms. Both belong to the fastest and most scalable family of algorithms in the literature with a linear computational complexity of $O(E)$. However, while the performance of LPA rapidly degrades for large mixing parameters, FluidC is able to produce relevant communities at all mixing parameters. The next algorithm in terms of scalability is the Multilevel algorithm, which takes roughly 5x more seconds to compute, and which scales slightly worse (see Figure \ref{fig:times} and the mentioned slopes). Thus, we consider FluidC to be highly recommendable for computing graphs of arbitrary large size.

In terms of diversity, FluidC is the first propagation-based algorithm to report competitive results on the LFR benchmark, and also the first propagation-based algorithm which can find a variable number of communities on a given graph. Providing coherent and diverse communities is particularly important for unsupervised learning tasks, such as CD, where typically there is not a single correct answer. To measure the diversity provided by FluidC, we computed its behaviour on multi-ground truth graphs, and compare it with the alternatives. Results indicate that FluidC uncovers sets of communities which may be consistently different than the ones obtained by the other algorithms. Considering both performance on LFR and diversity, we conclude that a thorough community analysis of a given graph would benefit from the inclusion of Multilevel, Infomap and FluidC.

\section*{Acknowledgements}
This work is partially supported by the Joint Study Agreement no. W156463 under the IBM/BSC Deep Learning Center agreement, by the Spanish Government through Programa Severo Ochoa (SEV-2015-0493), by the Spanish Ministry of Science and Technology through TIN2015-65316-P project and by the Generalitat de Catalunya (contracts 2014-SGR-1051), and by the Japan JST-CREST program.

\bibliographystyle{spbasic}
\bibliography{bibliography}

\begin{thebibliography}{16}
\providecommand{\natexlab}[1]{#1}
\providecommand{\url}[1]{{#1}}
\providecommand{\urlprefix}{URL }
\expandafter\ifx\csname urlstyle\endcsname\relax
  \providecommand{\doi}[1]{DOI~\discretionary{}{}{}#1}\else
  \providecommand{\doi}{DOI~\discretionary{}{}{}\begingroup
  \urlstyle{rm}\Url}\fi
\providecommand{\eprint}[2][]{\url{#2}}

\bibitem[{Blondel et~al(2008)Blondel, Guillaume, Lambiotte, and
  Lefebvre}]{blondel2008fast}
Blondel VD, Guillaume JL, Lambiotte R, Lefebvre E (2008) Fast unfolding of
  communities in large networks. Journal of statistical mechanics: theory and
  experiment 2008(10):P10,008

\bibitem[{Clauset et~al(2004)Clauset, Newman, and Moore}]{clauset2004finding}
Clauset A, Newman ME, Moore C (2004) Finding community structure in very large
  networks. Physical review E 70(6):066,111

\bibitem[{Fortunato(2010)}]{fortunato2010community}
Fortunato S (2010) Community detection in graphs. Physics reports
  486(3):75--174

\bibitem[{Girvan and Newman(2002)}]{girvan2002community}
Girvan M, Newman ME (2002) Community structure in social and biological
  networks. Proceedings of the national academy of sciences 99(12):7821--7826

\bibitem[{Lancichinetti and Fortunato(2009)}]{lancichinetti2009community}
Lancichinetti A, Fortunato S (2009) Community detection algorithms: a
  comparative analysis. Physical review E 80(5):056,117

\bibitem[{Lancichinetti et~al(2008)Lancichinetti, Fortunato, and
  Radicchi}]{lancichinetti2008benchmark}
Lancichinetti A, Fortunato S, Radicchi F (2008) Benchmark graphs for testing
  community detection algorithms. Physical review E 78(4):046,110

\bibitem[{Meusel et~al(2014)Meusel, Vigna, Lehmberg, and
  Bizer}]{meusel2014graph}
Meusel R, Vigna S, Lehmberg O, Bizer C (2014) Graph structure in the
  web---revisited: a trick of the heavy tail. In: Proceedings of the 23rd
  international conference on World Wide Web, ACM, pp 427--432

\bibitem[{Newman(2006)}]{newman2006finding}
Newman ME (2006) Finding community structure in networks using the eigenvectors
  of matrices. Physical review E 74(3):036,104

\bibitem[{Newman and Girvan(2004)}]{newman2004finding}
Newman ME, Girvan M (2004) Finding and evaluating community structure in
  networks. Physical review E 69(2):026,113

\bibitem[{Pons and Latapy(2005)}]{pons2005computing}
Pons P, Latapy M (2005) Computing communities in large networks using random
  walks. In: International Symposium on Computer and Information Sciences,
  Springer, pp 284--293

\bibitem[{Raghavan et~al(2007)Raghavan, Albert, and Kumara}]{raghavan2007near}
Raghavan UN, Albert R, Kumara S (2007) Near linear time algorithm to detect
  community structures in large-scale networks. Physical review E 76(3):036,106

\bibitem[{Reichardt and Bornholdt(2006)}]{reichardt2006statistical}
Reichardt J, Bornholdt S (2006) Statistical mechanics of community detection.
  Physical Review E 74(1):016,110

\bibitem[{Ronhovde and Nussinov(2009)}]{ronhovde2009multiresolution}
Ronhovde P, Nussinov Z (2009) Multiresolution community detection for megascale
  networks by information-based replica correlations. Physical Review E
  80(1):016,109

\bibitem[{Rosvall and Bergstrom(2008)}]{rosvall2008maps}
Rosvall M, Bergstrom CT (2008) Maps of random walks on complex networks reveal
  community structure. Proceedings of the National Academy of Sciences
  105(4):1118--1123

\bibitem[{Rosvall et~al(2009)Rosvall, Axelsson, and Bergstrom}]{rosvall2009map}
Rosvall M, Axelsson D, Bergstrom CT (2009) The map equation. The European
  Physical Journal Special Topics 178(1):13--23

\bibitem[{Yang et~al(2016)Yang, Algesheimer, and Tessone}]{yang2016comparative}
Yang Z, Algesheimer R, Tessone CJ (2016) A comparative analysis of community
  detection algorithms on artificial networks. Scientific Reports 6

\end{thebibliography}

\end{document}